# Design of a Low-Cost Prototype Underwater Vehicle

Ahsan Tanveer[1,a] and S.M. Ahmad[1,b]

[1]Faculty of Mechanical Engineering, GIK Institute of Engineering Sciences and Technology, Topi 23640, Khyber Pakhtunkhwa, Pakistan

Email address: [a)] ahsantanveer3883@gmail.com and [b)] smahmad@giki.edu.pk

*Abstract*— **In this study, a small, inexpensive remotely driven underwater vehicle that can navigate in shallow water for the purpose of monitoring water quality and demonstrating vehicle control algorithms is presented. The vehicle is operated by an onboard micro-controller, and the sensor payload comprises a turbidity sensor for determining the quality of the water, a depth sensor, and a 9-axis inertial measurement unit. The developed vehicle is an open frame remotely operated vehicle (ROV) with a small footprint and a modular physical and electrical architecture. With a net weight of 1.6 kg, a maximum depth rating of 20 meters, and a development cost of around $80, the ROV frame is composed of polyvinyl chloride tubes and has a length of 0.35 meters. As a ground station, a dedicated laptop shows crucial vehicle data in real time and can send commands to the vehicle. Initial testing in the pool demonstrates that the vehicle is completely operational and effectively complies with pilot commands.**

I. INTRODUCTION

Unmanned underwater vehicles (UUVs), including autonomous underwater vehicles (AUVs) and remotely operated vehicles (ROVs), are needed for a variety of underwater research and monitoring operations nowadays. However, given the harsh and unforgiving underwater environment as well as the high cost of underwater equipment, the development of such systems is a challenging and expensive endeavor. Nonetheless, over the past ten years, there has been a growing tendency toward the development of low-cost, readily available academic solutions for training future marine researchers and students. Some of the recent efforts to introduce a cost effective underwater robotic platform include an open-source AUV called HAL-Urabo [1], a LEGO based ROV [2], an open-frame ROV developed at Cornell [3], and a fish-inspired AUV [4]. Similar efforts have also been made in Pakistan to develop an inexpensive ROV. Indigenous marine robotic systems have been successfully developed by [5], [6], [7], and [8]. The majority of these systems offer solutions that are cost effective, but it is still challenging to design a system that is useful in academic settings utilizing components that are widely accessible.

In this regard, we propose an inexpensive and simple-to-replicate ROV concept for condition monitoring in shallow waters. The developed ROV is a fully functional underwater vehicle that can move in calm water, analyze the underwater environment, and exhibit different control algorithms. Design objectives for the developed vehicle include: a) ability to track and maintain reference input in depth and heading, b) allow for implementation and demonstration of various vehicle control algorithms, and c) continuous water condition monitoring using on-board sensors. The vehicle is intended to feature three thrusters managing three degrees of freedom, namely Yaw, Heave, and Surge, in order to achieve the design goals. Since the majority of the components of the developed ROV are produced in-house, it is a more affordable alternative to costly off-the-shelf ROVs.

There are four sections in the article. The hardware system design for the ROV is covered in Section II. Software and electronics are discussed in Section III. In Section IV, the results are summarized and a cost analysis is provided. The article is concluded in Section V.

## II. HARDWARE DESIGN

The built vehicle is an open-frame ROV, and its specifications are listed in Table I along with a front view of the ROV model in Fig. 1. A single hull alongside a 0.35-meter-long frame makes up the ROV. The maximum depth rating for the vehicle is 20 meters, with a total weight of 1.6 kilograms. The thrusters are fastened to the frame, and the hull—a plastic container fixed to the top of the ROV—contains all the electronics. A depth sensor (MS5803-05ba) and a 9-DOF IMU (MPU9250) are both employed to operate the vehicle. The data collection and control are done via the on-board Arduino Nano microcontroller. Three in-house developed thrusters are employed for propulsion. Each thruster has its own motor driver (L298N), and the vehicle is powered by a wire attached to a DC power source located at a ground station. Moreover, a turbidity sensor is used for water condition monitoring.

TABLE I. ROV SPECIFICATIONS

| Parameters | Specifications |
|---|---|
| Dimensions of the vehicle | 0.35 x 0.26 x 0.23 meters |
| Weight (in air) | 1.6 kg |
| Operating depth | Up to 20 meters |
| Endurance | Indefinite |
| Operating speed | 0.2-0.4 m/s |
| Actuators | Three 12-V DC motors |
| On-board computer | Arduino Nano |
| Sensor kit | MPU-9250, MS5083-05ba, Gravity Turbidity Sensor |
| Power and data transmission mechanism | DC wire (for power), CAT5 Ethernet cable (for data) |

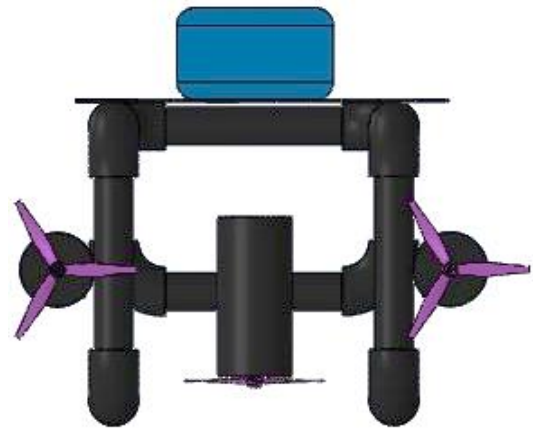

Fig. 1 Front view of the ROV CAD model.

### A. Buoyancy and Actuation

Due to shallow water constraints, floats are used in a relatively negatively buoyant configuration so that the vehicle body resurfaces in the absence of an input signal. A buoyant force of around 13 N was calculated by the buoyancy calculations.

Three 12V DC thrusters that are waterproofed by an epoxy coating placed on the motor power the ROV. Independent surge and yaw motion control is provided by two thrusters positioned on either side of the frame. The vehicle dives into the water as a result of being continuously propelled by the top thruster. By switching DC motors based on pulse width modulation (PWM), motor drivers change the speed of the motors.

## III. SOFTWARE AND ELECTRONICS ARCHITECTURE

The ROV uses a 9 DOF IMU, a depth transducer, and a turbidity detector as its sensors. The vehicle can monitor its yaw angle, depth, and water turbidity due to these sensors. The IMU uses the i2c interface to provide real-time data to the micro-controller and is made up of an accelerometer, a gyroscope, and a magnetic compass. Its primary function is to guarantee proper orientation of the ROV during field testing. The turbidity of water is measured using a gravity turbidity sensor, which spews a laser beam into water and then measures its transmittance and scattering rate.

A compact yet powerful Arduino Nano micro-controller controls the ROV. It acts as the vehicle's brain and gathers information from the IMU, depth sensor, and turbidity sensor before sending it back to the ground station PC. IMU feedback is used to regulate the yaw motion of the ROV. IMU information is obtained by Arduino, which then generates PWM to operate the motors after applying a complementary filter. Through a tether, power and control signals are sent to the vehicle from a ground-station PC, where the C++ program for the onboard computer is written. A PI controller is included in the onboard computer program, and the user may manually adjust its parameters using the Arduino programming environment.

## IV. RESULTS AND DISCUSSION

A shallow irrigation network, up to 20 metres deep, with high monitoring costs, can been regarded a good candidate for the use of the developed ROV. Since the power is provided from outside, the ROV has virtually endless endurance. Fig. 2 provide sample data for captured during pool tests. The bill of materials for the affordable ROV is shown in Table II. As can be seen, sensors and frame make up a significant portion of the cost. Generally speaking, expensive thrusters are the main reason ROV are expensive [9]. However, in this study, the indigenously developed version, which might not be highly efficient, allowed for a reduction in thruster costs. For the development of the ROV, around 80 USD were expended. The solution is helpful for low-cost academic use even if it may not be the best option for industrial implementation.

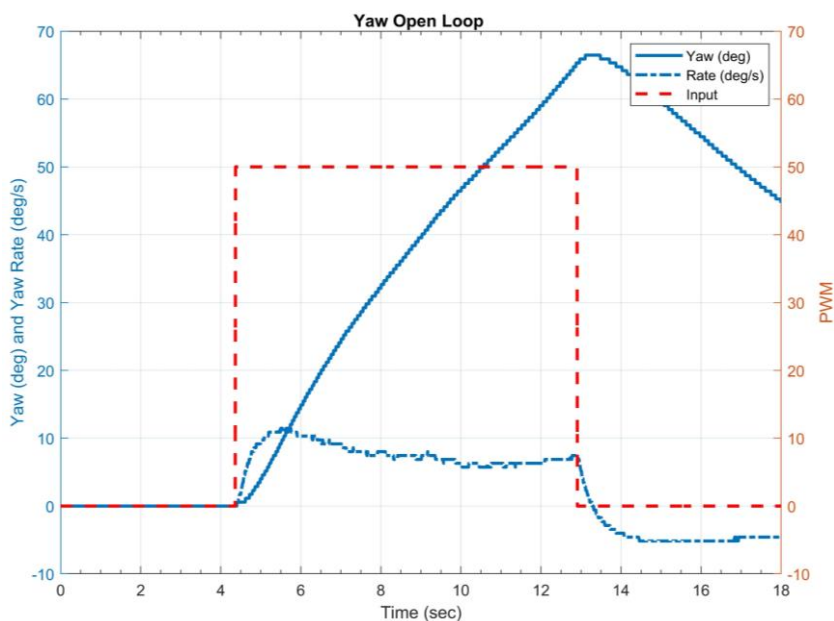

Fig. 2 Open-loop yaw response of the ROV obtained real-time in pool setting.

TABLE II.  ROV BILL OF MATERIALS

| Material/Equipment | Cost ($) |
|---|---|
| PVC frame | 15 |
| Motor thrusters | 14 |
| Motor drivers | 4 |
| Micro-controller | 8 |
| Sensor kit | 25 |
| Wiring and communication | 4 |
| Miscelleneous | 10 |
| **Total** | **USD 80** |

## V. CONCLUSION

This study presents the design and development of a domestically manufactured ROV. The design of a prototype underwater vehicle for demonstration and implementation of control algorithms is the main goal of this work. As a result of the integrated subsystems working together, ROV's subsystems are in perfect synchronization and communication with one another. Initial testing in the pool demonstrates that the vehicle is completely operational and effectively complies with pilot commands. Additionally, the ground control center receives real-time logs of all critical vehicle states. In sum, the developed ROV is appropriate for shallow underwater applications because to its sturdy construction, dependable water tightening mechanism, and flawless communication system.